\documentclass[aps,pre,reprint,10pt,groupedaddress]{revtex4-1}
\usepackage{amssymb,amsfonts,amsmath}
\usepackage[dvips]{graphicx}
\usepackage{epsfig}

\DeclareMathOperator{\sech}{sech}

\begin{document}

\title{Reentrance in an active spin glass model}

\author{Kevin R. Pilkiewicz}\affiliation{Department of Chemistry and Biochemistry, University of Colorado, Boulder}
\author{Joel D. Eaves}\email{joel.eaves@colorado.edu}\affiliation{Department of Chemistry and Biochemistry, University of Colorado, Boulder}

\date{\today}

\begin{abstract}
Active matter, whose motion is driven, and glasses, whose dynamics are arrested, seem to lie at opposite ends of the spectrum in nonequilibrium systems. In spite of this, both classes of systems exhibit a multitude of stable states that are dynamically isolated from one another. While this defining characteristic is held in common, its origin is different in each case: for active systems, the irreversible driving forces can produce dynamically frozen states, while glassy systems vitrify when they get kinetically trapped on a rugged free energy landscape. In a mixture of active and glassy particles, the interplay between these two tendencies leads to novel phenomenology. We demonstrate this with a spin glass model that we generalize to include an active component. In the absence of a ferromagnetic bias, we find that the spin glass transition temperature depresses with the active fraction, consistent with what has been observed for fully active glassy systems. When a bias does exist, however, a new type of transition becomes possible: the system can be cooled out of the glassy phase. This unusual phenomenon, known as reentrance, has been observed before in a limited number of colloidal and micellar systems, but it has not yet been observed in active glass mixtures. Using low order perturbation theory, we study the origin of this reentrance and, based on the physical picture that results, suggest how our predictions might be measured experimentally.
\end{abstract}

\maketitle

\section{Introduction}

Active systems, those whose particles exhibit externally driven or self-propelled motion, challenge standard descriptions of matter. While a growing amount of evidence from both simulations and experiments suggests that the dynamical structural transitions observed in active systems bear more than just a superficial resemblance to the thermodynamic phase transitions of systems at equilibrium,\cite{EGeigant,Evans,Vicsek,Tu,Chate,Larralde,Hagan1,SolonTailleur,PanicModel} the microscopically irreversible dynamics that drive these transitions can lead to a state space comprised of many similar steady-state configurations that are dynamically estranged from one another. This sort of configurational landscape is also observed in glasses, though there it is achieved through a different mechanism: kinetic trapping on a corrugated free energy surface.

In this paper we study a mixture of active and glassy particles to probe what transpires when these disparate mechanisms compete and interact. Fig. 1 illustrates how such a mixture differs from a glassy system in which all the particles are active. In a fully active system, every particle will have its average energy increased by the external driving, making it possible, in many cases, to map the active system onto its passive counterpart through an effective temperature (Fig. 1({\bf a})).\cite{LoiMossa,DJDurian,Wolynes1,BerthKurch} This suggests that so long as the driving is not too excessive, the same thermodynamic phases will be observed, only at lower temperatures (or higher densities), a conclusion that has been borne out in a number of studies.\cite{BerthKurch,Zipp,Lowen,MDijkstra}

When only a fraction of the system is active, however, the uneven distribution of energy will stabilize some configurations of the system while destabilizing others (Fig. 1({\bf b})). This has the potential to radically alter the system's phase diagram and lead to new physical phenomena. Consistent with this expectation, fully active systems in which a fraction of the particles have a higher motility have been observed to exhibit novel patterns of phase separation.\cite{LGBrunnet,Hagan2}

The glass forming system we study in this paper is a generalization of the mean field Ising spin glass, also known as the Sherrington-Kirkpatrick (SK) model. In the appropriate limits, fractional activation can be approximated as fractional annealing, and we show how this annealing modifies quantities like the free energy and the magnetization. After examining how these modifications alter the familiar SK phase diagram, we demonstrate that while some phase boundaries on the diagram merely shift or elongate, others change more drastically and allow transition pathways between phases that were not possible in the fully quenched model. Most notable among these is a reentrant spin glass transition\cite{footnote1} in which the spin glass can be cooled into a ferromagnet and then back to a spin glass as the temperature is lowered at fixed ferromagnetic bias. This behavior is similar in character to what has been observed in some colloid and micellar systems\cite{TEckert,WPoon,WRChen,JGrandjean} as well as numerous simulated systems.\cite{KDawson,CMishra,LowenLikos,MarkReich} A perturbation theory argument reveals the physical origin of this phenomenon, and we show, to leading order, that the effect of the fractional activation on the free energy landscape is consistent with the physical picture in Fig. 1({\bf b}). The fundamental mechanisms uncovered by this analysis extend beyond the specific magnetic interactions studied here, and we conclude with a discussion of how our results can be generalized to more complex systems and how such a system might be studied experimentally.

\begin{figure}[ht!]
\includegraphics[width=8.5cm,height=8.5cm,keepaspectratio=true]{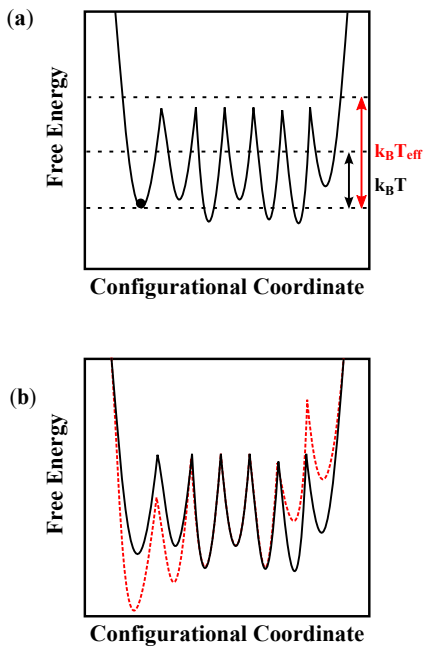}
\caption{Full versus fractional activation. ({\bf a}) A schematic representation of the free energy landscape for a glassy system. The black dot denotes a reference low energy configuration of the system, and the black dashed lines delineate the amount of thermal energy ($k_BT$) available to the system before activation and the larger amount of effective thermal energy ($k_BT_{eff}$) available afterwards due to the driving of the system. Prior to activation, the system is kinetically trapped in the indicated minimum, but activation postpones this trapping to lower temperatures. ({\bf b}) The same free energy landscape before (solid black curve) and after (dashed red curve) fractional activation of the system. The uneven distribution of energy in the system stabilizes some states while destabilizing others.}
\end{figure}

\section{The Model}

We start with the standard Sherrington-Kirkpatrick model, which consists of $N$ Ising spins interacting according to the following Hamiltonian.
\begin{equation}\label{SK_Hamiltonian}
H=-\sum_{(ij)}J_{ij}S_iS_j-h\sum_{i}S_i
\end{equation} 
In the above, the Ising spin variables $S_i$ can only take values of $+1$ or $-1$, $h$ is an external magnetic field, and the first sum is over all $N(N-1)/2$ distinct pairs of spins. The coupling constants $J_{ij}$ are chosen from a Gaussian distribution.
\begin{equation}\label{Jij_Dist}
P(J_{ij})=\left(\frac{N}{2\pi J^2}\right)^{1/2}\exp\left[\frac{-N(J_{ij}-J_0/N)^2}{2J^2}\right]
\end{equation}

The usual motivation for these random couplings is that the strength and sign of the exchange interaction, $J_{ij}$, varies as a function of the distance between each pair of spins. In a disordered material, these distances will be stochastic, so, for a sufficiently large system, one can approximately select the coupling constants for these interactions from the distribution in equation \eqref{Jij_Dist}. For a given sample, the $J_{ij}$ do not change and thus are considered ``quenched'' interactions, but when the free energy of the whole system is computed, it must be averaged over all realizations of the coupling constants.

We generalize this model to allow for a fixed fraction, $\mu$, of the spins to become active. We imagine there is an external driving force coupled to these spins, as well as a frictive force that keeps the system in a steady state. In the limit of strong activation, the exchange couplings of the active spins will fluctuate on time scales that are short compared to the spin relaxation time of the quenched degrees of freedom, so the steady state of this fractionally active system may be approximated by the thermal equilibrium of a fractionally annealed system. We term the resulting model the ``fractionally annealed Sherrington-Kirkpatrick'' (FASK) model. A pictorial representation of this model is shown in Fig. 2({\bf a}). Fig. 2({\bf b}) emphasizes the basic similarities between our model and a more realistic fractionally active glass former, discussed at the end of the paper.

\begin{figure}[ht!]
\includegraphics[width=8.5cm,height=8.5cm,keepaspectratio=true]{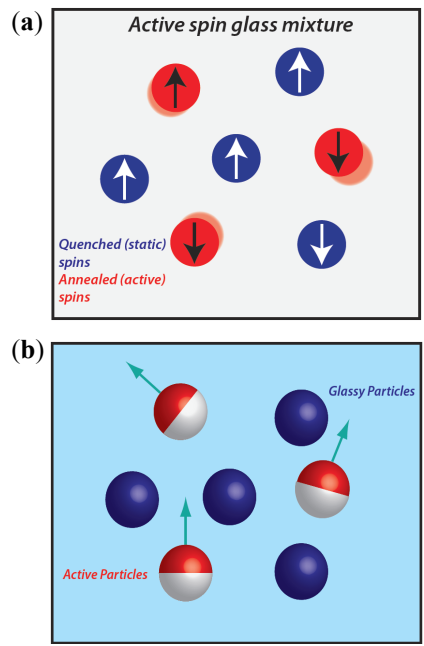}
\caption{Depiction of the model. ({\bf a}) A simple pictorial representation of the FASK model, with the quenched, inactive spins drawn as blue circles and the active spins drawn as red circles. The arrows inside each circle indicate the particle's spin state, and the motion of the active spins is depicted as motion blur. ({\bf b}) A pictorial representation of a potential experimental system that would behave as a fractionally active glass former. The blue spheres represent silica beads, the half red, half white spheres represent silica beads that are half coated in platinum, and the light blue background represents hydrogen peroxide solvent. Arrows indicate the direction of self-propulsion for the active colloid particles.}
\end{figure} 

To construct the partition function for this system, we first divide the pairs of spins into two non-intersecting sets: a set with annealed interactions $\mathcal{A}\equiv\{(ij)\,\,\vert\,\, i=1,...,\mu N,\,\, j>i\}$ and a set with quenched interactions $\mathcal{Q}\equiv\{(ij)\,\,\vert\,\, i=\mu N +1,...,N,\,\, j>i\}$, where we have numbered the active spins with labels $1$ through $\mu N$, and the passive spins with labels $\mu N+1$ through $N$. This division allows us to rewrite the first sum on the right of equation \eqref{SK_Hamiltonian} as
\begin{equation*}
\sum_{(ij)}J_{ij}S_iS_j=\sum_{(ij)\in\mathcal{A}}J_{ij}S_iS_j+
\sum_{(ij)\in\mathcal{Q}}J_{ij}S_iS_j
\end{equation*} 
A similar factorization of the product over $(ij)$ allows us to write down the desired partition function.
\begin{align*}
Z_{\mu}&=tr_{S}\left\{\int\prod_{(ij)\in\mathcal{A}}
\left(\frac{N^{1/2}dJ_{ij}}{(2\pi J^2)^{1/2}}\right)\right. \nonumber \\
&\exp\left[\sum_{(ij)\in\mathcal{A}}
\left(\beta J_{ij}S_iS_j-\frac{N(J_{ij}-J_0/N)^2}{2J^2}\right)\right] \nonumber \\
&\left.\times\exp\left[\sum_{(ij)\in\mathcal{Q}}\beta J_{ij}S_iS_j+\beta h\sum_{i}S_i\right]\right\}
\end{align*} 
The trace in this expression is over the $2^N$ distinct spin configurations of the system. After performing the Gaussian integrals over the annealed interactions, this partition function can be reduced to the following form.
\begin{align*}
Z_{\mu}&=\exp\left[\left(\frac{\beta J}{2}\right)^2\mu(2-\mu)N\right] \nonumber \\
&\times tr_S\exp\left[\beta\sum_{(ij)\in\mathcal{Q}}(J_{ij}-J_0/N)S_iS_j\right. \nonumber \\
&+\left.\beta (J_0/N)\sum_{(ij)}S_iS_j+\beta h\sum_{i}S_i\right]
\end{align*}
It is important to note that although we are treating this active system as if it were at thermal equilibrium, for $\mu>0$ we are still driving it far from the equilibrium of the fully quenched model.

The Helmholtz free energy per spin, $f$, in the FASK model, averaged over the quenched interactions, can be computed using the usual replica trick.\cite{CastellCavagna,FischerHertz}
Since we will primarily be concerned with phase boundaries, it is sufficient to evaluate the free energy within the assumption of replica symmetry. The derivation proceeds similarly to that of the standard SK model free energy,\cite{FischerHertz,SKModel} so only the final result will be shown here.
\begin{align}\label{FASK_f}
-\beta f&=\left(\frac{\beta J}{2}\right)^2\left[(1-q_{\mu})^2+2\mu q_{\mu}\right] -\frac{\beta J_0}{2}M^2\nonumber \\
&+\frac{1-\mu}{(2\pi)^{1/2}}\int_{-\infty}^{\infty}dz\,e^{-\frac{1}{2}z^2}
\ln\left[2\cosh\eta(z)\right] \nonumber \\
&+\mu\ln\left[2\cosh\beta
\left(J_0M+h\right)\right]
\end{align}
In the above, $\eta(z)=\beta\left(Jq_{\mu}^{1/2}z+J_0M+h\right)$ and the order parameters $q_{\mu}$ and $M$ are defined through the following self-consistency relations.
\begin{equation*}
q_{\mu}=\frac{1-\mu}{(2\pi)^{1/2}}\int_{-\infty}^{\infty}dz\,
e^{-\frac{1}{2}z^2}\tanh^{2}\eta(z)
\end{equation*}
\begin{align}\label{OrderParam_M}
M&=\frac{1-\mu}{(2\pi)^{1/2}}\int_{-\infty}^{\infty}dz\,
e^{-\frac{1}{2}z^2}\tanh\eta(z) \nonumber \\
&+\mu\tanh\beta(J_0M+h)
\end{align}
In the extreme cases of $\mu=0$ and $\mu=1$, equation \eqref{FASK_f} reduces, as required, to the familiar results of the fully quenched SK model and the fully annealed mean field Ising model, respectively.

Differentiating equation \eqref{OrderParam_M} with respect to the field $h$ and taking the limit $h\rightarrow 0$ leads to an expression for the zero field magnetic susceptibility.
\begin{equation}\label{Suscept}
\chi_M=\frac{1-q}{k_BT-J_0(1-q)}
\end{equation} 
This expression is identical to that obtained for the usual SK model, except that now the overlap order parameter $q$ is defined as follows.
\begin{equation*}
q=q_{\mu}+\mu\tanh^{2}\beta(J_0M)
\end{equation*}

These results are all for the replica symmetric solution of the free energy. The validity of this solution is determined by the following stability condition, which is analogous to that found by de Almeida and Thouless\cite{AT} for the SK model.
\begin{equation}\label{StabilityCond}
\frac{(\beta J)^2(1-\mu)}{(2\pi)^{1/2}}\int_{-\infty}^{\infty}dz\,e^{-\frac{1}{2}z^2}
\sech^4\eta(z)<1
\end{equation}
A detailed derivation of equations \eqref{FASK_f}--\eqref{StabilityCond} may be found elsewhere.\cite{FASK2}

\section{Results}

For convenience, we will use reduced units for the remainder of the paper where temperature is scaled by $k_B/J$ and all energies are given in units of $J$.

In the $T$-$h$ plane, there is a single phase transition occurring at $h=0$ between a paramagnetic phase ($q=0$, $M=0$) and a spin glass phase ($q\neq 0$, $M=0$). The spin glass transition temperature, $T_f$, can be computed as a function of the active fraction $\mu$ by finding the temperature at which equation \eqref{StabilityCond} becomes an equality for $q_{\mu}$, $J_0$, and $h$ all set to zero. The result is plotted in Fig. 3({\bf a}).
\begin{equation*}
T_f=\sqrt{1-\mu}
\end{equation*}
In their treatment of a fully active spin glass system, Berthier and Kurchan\cite{BerthKurch} found a roughly linear relationship between the magnitude of their driving force and the depression of their glass transition temperature, and though our result becomes highly nonlinear as $\mu$ approaches unity, for $\mu$ less than roughly $0.5$, a linear fit is very good (see Fig. 3({\bf a})). For small to moderate amounts of activation, the shift of the paramagnetic to spin glass transition temperature is qualitatively similar regardless of whether the whole system gets partially annealed or one fraction of it gets fully annealed.   

\begin{figure}[ht!]
\includegraphics[width=8.5cm,height=4.25cm,keepaspectratio=true]{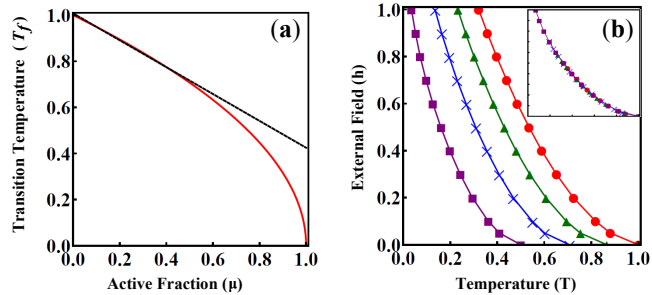}
\caption{Phase diagram in the $T-h$ plane. ({\bf a}) The spin glass transition temperature $T_f$ plotted versus the active fraction $\mu$. The black line represents a best fit for the curve up to $\mu=0.5$. The slope of this line is roughly $0.58$, a little larger than what one would get from a linear Taylor expansion about $\mu=0$. ({\bf b}) The Almeida-Thouless stability line in the $T$-$h$ plane, plotted for active fractions $\mu=0$ (red), $\mu=0.25$ (green), $\mu=0.50$ (blue), and $\mu=0.75$ (purple). The inset shows that these curves all collapse onto the $\mu=0$ master curve when the temperature and field are both scaled by a factor of $\left(1-\mu\right)^{-1/2}$.}
\end{figure} 

We can go further and use the stability condition of equation \eqref{StabilityCond} to plot the entire Almeida-Thouless (AT) stability line for different values of $\mu$. The results are shown in Fig. 3({\bf b}). While the entire curve is shifted to lower temperatures with increasing active fraction, the amount each point gets shifted decreases with increasing field due to all the curves converging towards infinite field as $T\rightarrow 0$. If one scales the temperature by a factor of one over $T_f$, it is clear that each of these curves will cross the temperature axis at $T=1$, but, interestingly, if the external field is also scaled by that same factor, the curves for different $\mu$ all collapse onto the fully quenched curve (see the inset of Fig. 3({\bf b})).  

The FASK model phase diagram is much richer in the $J_0$-$T$ plane, because in addition to a paramagnetic phase and a spin glass phase with $M=0$, there is also a ferromagnetic phase and a spin glass phase with $M\neq 0$, often referred to as a mixed phase.\cite{Toulouse1} The boundary between the region of the phase diagram with $M=0$ and that with $M\neq 0$ can be determined by finding where the susceptibility diverges. Using equation \eqref{Suscept}, one finds that the Curie temperature $T_c$ is given as a function of $J_0$ by the following relation.
\begin{equation*}
T_c=J_0(1-q_{\mu}(T_c))
\end{equation*}
Note that when $q_{\mu}=0$ (in the paramagnetic phase), the above simplifies to $T_c=J_0$. The remaining phase boundaries can be found by using the stability condition of equation \eqref{StabilityCond}. An example of the phase diagram that results from these considerations is shown in Fig. 4({\bf a}), for $\mu=0.50$. The replica symmetric phase diagram of the fully quenched model is plotted in light gray for comparison.

\begin{figure}[ht!]
\includegraphics[width=8.5cm,height=8.5cm,keepaspectratio=true]{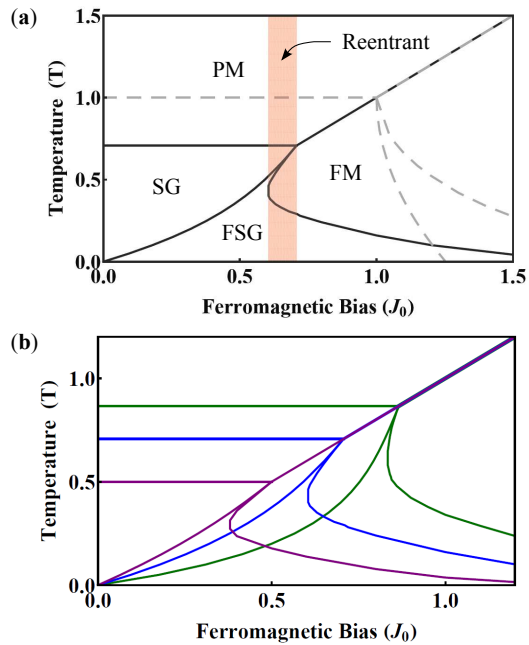}
\caption{Phase diagram in the $J_0$-$T$ plane. ({\bf a}) The FASK model phase diagram plotted in the $J_0$-$T$ plane for $\mu=0.5$. The dashed gray lines are the phase curves for the fully quenched SK model ($\mu=0$). The labels PM, FM, SG, and FSG refer to the paramagnetic, ferromagnetic, spin glass, and ferromagnetic spin glass phases respectively. The shaded red region gives the range of $J_0$ for which reentrance is possible. ({\bf b}) Plots of the FASK model phase diagram for $\mu=0.25$ (green), $\mu=0.50$ (blue), and $\mu=0.75$ (purple). In each phase diagram, the horizontal line is given by $T_f=\sqrt{1-\mu}$ (see Fig. 3({\bf a})). As $\mu$ approaches one, the curve separating the two spin glass phases approaches the line $T=J_0$, and the region where reentrance can occur increases in size.}
\end{figure} 

A non-zero active fraction causes the paramagnetic/spin glass transition line to shift to a lower temperature $T=\sqrt{1-\mu}$ and terminate at a lower value of $J_0$, also equal to $\sqrt{1-\mu}$. The paramagnetic/ferromagnetic transition line is still the curve $T=J_0$, but now it terminates at the point $\left(\sqrt{1-\mu},\sqrt{1-\mu}\right)$ instead of $(1,1)$. The paramagnetic phase consequently occupies a trapezoidal region of the phase diagram for all $\mu<1$, whose area grows linearly with $\mu$. Specifically, for an increase in active fraction equal to $\Delta\mu$, this area changes by $(1/2)\Delta\mu$. 

The impact of a non-zero active fraction on the low temperature region of the phase diagram is more dramatic. For $\mu>0$, the spin glass/ferromagnetic spin glass phase boundary bends in the opposite direction, connecting the points $\left(\sqrt{1-\mu},\sqrt{1-\mu}\right)$ and $(0,0)$. This means that at sufficiently low temperatures, the system will become partially ordered for all $J_0>0$. Though the replica symmetric solution is not valid in this region of the phase diagram, the phase boundary it predicts does approach the line $T=J_0$ as $\mu\rightarrow 1$, as physically required, so it is likely to be at least qualitatively correct. The AT line separating the ferromagnetic and ferromagnetic spin glass phases also changes shape for $\mu>0$, bending in on itself to create a reentrant region where it is possible, just by lowering the temperature, for the system to transition from a spin glass to a ferromagnet back to a spin glass.

In most systems with a reentrant glass transition, repulsive interactions dominate the higher temperature glass phase while attractive interactions dominate in the lower temperature glass.\cite{TEckert,WPoon,WRChen,JGrandjean} If, as in a lattice gas,\cite{KHuang} one views antiferromagnetism and ferromagnetism as repulsion and attraction, respectively, then the same basic phenomenology holds in the FASK model. The initial spin glass formation is driven by antiferromagnetic interactions that compete with the ferromagnetic bias to cause frustration, while the reentrant spin glass is characterized by some degree of ferromagnetic order, a result of the more prevalent interactions winning out at low temperature. The range of $J_0$ over which reentrance can occur is shown as a shaded region in Fig. 4({\bf a}), and, in Fig. 4({\bf b}), a side-by-side plot of the FASK model phase diagram for several values of $\mu$ reveals that this range grows with increasing active fraction.

We can better understand the origin of reentrance in this model by performing a perturbative analysis of the magnetization, similar to that used to derive the Born approximation in quantum mechanics. Equation \eqref{OrderParam_M} can be rewritten as $M=(1-\mu)M_q+\mu M_a$, where $M_q$ is the expression for the magnetization of a fully quenched system ($\mu=0$) and $M_a$ is the corresponding expression for a fully annealed system ($\mu=1$). The mobile and immobile spins both contribute to the total magnetization proportionally to their fractional composition of the system, though these contributions are coupled by their mutual dependence on the same total magnetization $M$. If we were to ignore this coupling, a zeroth order approximation to the total magnetization would be
\begin{equation}\label{ZerothOrder}
M(J_0,T)\approx(1-\mu)M_{q}^{*}(J_0,T)+\mu M_{a}^{*}(J_0,T),
\end{equation}
where $M_{q}^{*}$ and $M_{a}^{*}$ are the magnetizations that a pure quenched and pure annealed system would have, respectively, at the given values of $J_0$ and $T$.

Inserting the zeroth order solution back into the right hand side of equation \eqref{OrderParam_M} for $h=0$, one obtains the following result.
\begin{align}\label{M_Coupling}
M&\approx\frac{1-\mu}{(2\pi)^{1/2}}\int_{-\infty}^{\infty}dz\,
e^{-\frac{1}{2}z^2}\tanh\left[\frac{q^{1/2}_{\mu}z+J_0M_{q}^{*}
+\mu h_{eff}}{T}\right] \nonumber \\
&+\mu\tanh\left[\frac{J_0M_{a}^{*}-(1-\mu) h_{eff}}{T}\right]
\end{align}
In the above, we have defined an effective magnetic field as follows.
\begin{equation}\label{heff}
h_{eff}\equiv J_0(M_{a}^{*}-M_{q}^{*}) 
\end{equation}
The interpretation of this result is as follows. Reentrance is only observed when $J_0<1$ and $T<J_0$, in which case $M_{q}^{*}=0$ and $M_{a}^{*}\neq 0$. The inactive component of the system thus feels, to leading order, an effective magnetic field from the active component that can cause it to align out of the spin glass phase into a ferromagnet. The fact that $h_{eff}$ is proportional to $\mu$ in the first term on the right hand side of equation \eqref{M_Coupling} also explains why increasing the active fraction broadens the range of $J_0$ over which reentrance occurs.

If the system is fully annealed, we can recast the Hamiltonian using the Weiss form of mean field theory.\cite{WeissMFT}
\begin{equation*}
H=-J_0\sum_{i=1}^{N}M_{a}^{*}S_i
\end{equation*}
In the above, we have neglected the term that comes from integrating over the annealed degrees of freedom, since at fixed $T$ and $\mu$ it is just a constant. Expanding about this solution by replacing $M_{a}^{*}$ with the zeroth order approximation in equation \eqref{ZerothOrder}, we get the following approximate result.
\begin{equation}\label{ExpandedH}
H\approx -\sum_{(ij)}J_{ij}S_i S_j-\mu h_{eff}\sum_{i=1}^{N}S_i
\end{equation} 
In the above, $h_{eff}$ is the same as in equation \eqref{heff}, and all coupling constants are quenched. Equation \eqref{ExpandedH} suggests that for $\mu$ close to unity, the system looks, to leading order, like a fully quenched system in the presence of an effective magnetic field. This field selectively stabilizes configurations of the system that have a net alignment with it and destabilizes those that align against it, consistent with the physical picture depicted in Fig. 1({\bf b}). The difference $M_{a}^{*}-M_{q}^{*}$ is largest for $T<J_0$ and $J_0<1$, which is precisely where the phase diagram of the FASK model differs most strikingly from that of the fully quenched system. 

\section{Discussion}

In the mean field Ising spin glass, activating a fraction of the system gives rise to new physical phenomena--most notably a reentrant transition from the spin glass phase to the ferromagnetic phase. The origin of this reentrant behavior lies in the fact that the active component will start to magnetize at low temperatures, generating a local magnetic field that can, for a certain range of $J_0$, overpower the frustrated interactions of the passive spins and induce a net magnetization in the entire system. 

One can imagine reentrance occurring in other systems through a parallel mechanism. In a fractionally activated glass forming liquid, for example, the active particles would be harder to vitrify than the passive particles, leading to a glass phase with pockets of active particles in a liquid-like state. It is conceivable that for a limited range of densities, these pockets could transfer enough of their driven energy to the surrounding passive particles to break them out of their cages and cause reentrance to the liquid phase. This effect will be enhanced if an aligning mechanism is present, in which case the active particles will tend to exhibit cooperative motion.

While it is easy enough in theoretical treatments to leave the task of selectively activating a fraction of the system to some Maxwellian mephisto, designing a practical experimental method for accomplishing this task is more difficult. Recent experimental work has shown that silica particles, a well-known colloidal glass former,\cite{silica1,silica2} half coated in platinum undergo self-propelled motion in hydrogen peroxide.\cite{silica3} A dense colloidal suspension of silica particles in which only a fraction were so coated might therefore be viable as a fractionally active glass forming system (see Fig. 2({\bf b})). 

Simple spin glass models have led to many insights into the nature of the glassy state, and these concepts and tools have had applications in fields as distinct as protein folding and neurosicence. In the nascent field of active glass formers, spin glass models will likely continue to play a key role.


\begin{thebibliography}{10}
\bibitem{footnote1}
Note that the ferromagnetic spin glass (or mixed) phase is referred to as a reentrant phase in some spin glass literature\cite{AIto,NHur}, but this is not the sense in which we use the term.
\end{thebibliography}
\end{document}